\documentclass[prl,twocolumn,superscriptaddress,tightenlines,floatfix,showpacs,a4paper]{revtex4}
\usepackage{amsfonts}
\usepackage{amsmath,amsthm}
\usepackage{amssymb}
\usepackage{graphicx}
\usepackage{amsmath}
\usepackage{color}

\providecommand{\U}[1]{\protect\rule{.1in}{.1in}}
\newtheorem{theorem}{Theorem}

\newtheorem{lemma}[theorem]{Lemma}

\theoremstyle{definition}

\renewenvironment{proof}[1][Proof]{\noindent\textbf{#1.} }{\ \rule{0.5em}{0.5em}}

\DeclareMathOperator{\sgn}{sgn}

\begin{document}
\title{Preparation contextuality powers parity-oblivious multiplexing}
\author{Robert W. Spekkens}
\affiliation{DAMTP, University of Cambridge, Cambridge, United Kingdom CB3 0WA}
\author{D. H. Buzacott}
\affiliation{Centre for Quantum Computer Technology, Griffith University, Brisbane 4111, Australia}
\affiliation{Centre for Quantum Dynamics, Griffith University, Brisbane 4111, Australia}
\author{A. J. Keehn}
\affiliation{Centre for Quantum Computer Technology, Griffith University, Brisbane 4111, Australia}
\affiliation{Centre for Quantum Dynamics, Griffith University, Brisbane 4111, Australia}
\author{Ben Toner}
\affiliation{Centrum voor Wiskunde en Informatica, Kruislaan 413, 1098 SJ Amsterdam, The Netherlands}
\author{G. J. Pryde}
\affiliation{Centre for Quantum Computer Technology, Griffith
University, Brisbane 4111, Australia}
\affiliation{Centre for
Quantum Dynamics, Griffith University, Brisbane 4111, Australia}

\begin{abstract}
In a noncontextual hidden variable model of quantum theory, hidden variables determine the outcomes of every
measurement in a manner that is independent of how the measurement is implemented. Using a generalization of
this notion to arbitrary operational theories and to preparation procedures, we demonstrate that a particular
two-party information-processing task, ``parity-oblivious multiplexing,'' is powered by contextuality in the
sense that there is a limit to how well any theory described by a noncontextual hidden variable model can
perform. This bound constitutes a ``noncontextuality inequality'' that is violated by quantum theory. We report
an experimental violation of this inequality in good agreement with the quantum predictions. The experimental
results also provide the first demonstration of 2-to-1 and 3-to-1 quantum random access codes.

\end{abstract}
\maketitle

The Bell-Kochen-Specker theorem \cite{BKS} shows that the predictions of quantum theory are inconsistent with a
hidden variable model having the following feature: if $A$, $B$ and $C$ are Hermitian operators such that $A$
and $B$ commute, $A$ and $C$ commute, but $ B$ and $C$ do not commute, then the value predicted to occur in a
measurement of $A$ does not depend on whether $B$ or $C$ was measured simultaneously.  This feature is called
``noncontextuality.'' Significantly, it is only well-defined for models of quantum theory (and then only for
projective measurements and deterministic models) \cite{Spe05}. By contrast, Bell's definition of a \emph{local}
model applies to any theory that can be described operationally \cite{Bell64}. Consequently, whereas one can
test whether or not experimental statistics are consistent with a local model (by testing whether or not they
satisfy Bell inequalities), there is no way to test whether or not experimental statistics are consistent with a
noncontextual model (and no way of defining associated ``noncontextuality inequalities'') \emph{unless} one
generalizes the traditional notion of noncontextuality in such a way that it makes no reference to the quantum
formalism. Suggestions for such a formulation have been made by several authors~\cite{previouswork}. A
particularly natural generalization (and slight modification) which applies to all models (deterministic or not)
of any operational theory has been proposed in Ref.~\cite{Spe05}. We here derive a noncontextuality (NC)
inequality based on this notion.

Because information-theoretic tasks can be characterized entirely in terms of experimental statistics, one can
explore whether theories that violate NC inequalities may provide information-theoretic advantages over theories
that satisfy these inequalities. \ We prove that this is indeed the case for a task which we call
\emph{parity-oblivious multiplexing}, a kind of two-party secure computation.  (The notion that contextuality
might yield an advantage for multiplexing tasks was first put forward by Galv\~{a}o \cite{Galvao}.) The NC
inequality we derive provides a bound on the probability of success in this task and we demonstrate a quantum
protocol for parity-oblivious multiplexing for which the probability of success exceeds the noncontextual bound.

Finally, we report an experimental implementation of this protocol that achieves a probability of success in
good agreement with the quantum result and in violation of the NC inequality.

\textbf{Operational theories and noncontextual models. }In an
operational theory, the primitives of description are preparations
and measurements, specified as instructions for what to do in the
laboratory. The theory simply provides an algorithm for calculating
the probability $p(k|P,M)$ of an outcome $k$ of measurement $M$
given a preparation $P.$ \ As an example, in quantum theory, every
preparation $P$ is represented by a density operator $\rho_{P},$
every measurement $M$ is represented by a positive operator valued
measure $\{E_{M,k}\},$ and the probability of outcome $k$ is given
by $p(k|P,M)=\textrm{Tr}\left( \rho_{P}E_{M,k}\right) .$

In a hidden variable model of an operational theory,
a preparation procedure is assumed to prepare a system with certain
properties and a measurement procedure is assumed to reveal
something about those properties. \ The set of all variables
describing the system is denoted $\lambda.$ It is presumed that for
every preparation $P$, there is a probability distribution
$p(\lambda|P)$ such that implementing $P$ causes the system to be
prepared in physical state $\lambda$ with probability
$p(\lambda|P).$ Similarly, it is presumed that for every measurement
$M$, there is a distribution $p(k|\lambda,M)$ such that implementing
$M$ on a system described by $\lambda$ yields outcome $k$ with
probability $p(k|\lambda,M).$ For the hidden variable model to
reproduce the predictions of the operational theory, it must satisfy
$p(k|P,M)=\int\mathrm{d}\lambda{p}(k|\lambda,M)p(\lambda |P).$

A hidden variable model is \emph{preparation noncontextual} if the
following implication holds
\begin{equation}
\forall M : p(k|P,M)=p(k|P^{\prime},M) \rightarrow
\;p(\lambda|P)=p(\lambda|P^{\prime}), \label{eq:PNC}%
\end{equation}
that is, if two preparations yield the same statistics for all possible
measurements then they are represented equivalently in the hidden variable
model. Similarly, \emph{measurement noncontextuality} is the condition that%
\begin{equation}
\forall P : p(k|P,M)=p(k|P,M^{\prime})\rightarrow\;p(k|\lambda ,M)=p(k|\lambda,M^{\prime}), \label{eq:MNC}
\end{equation}
that is, if two measurements have the same statistics for all
possible preparations then they are represented equivalently in the
model. \ More details can be found in Ref. \cite{Spe05}. \ An NC
inequality is any inequality on experimental statistics that follows
from the assumption that there exists a hidden variable model that
is preparation and measurement noncontextual. It is of the form
$f(p(k|P_{1},M_{1}),p(j|P_{2},M _{2}),...)\leq C$ for some function
$f$ and constant $C$.

\textbf{Parity-oblivious multiplexing. }Suppose that Alice and Bob wish to perform the following
information-processing task, which we call \emph{n-bit parity-oblivious multiplexing}. Alice has as input an
$n$-bit string $x$ chosen uniformly at random from $\{0,1\}^{n}$. Bob has as input an integer $y$ chosen
uniformly at random from $\{1,\ldots,n\}$ and must output the bit $b=x_{y},$ that is, the $y$th bit of Alice's
input. Alice can send a system to Bob encoding information about her input, however there is a cryptographic
constraint: no information about any parity of $x$ can be transmitted to Bob. More specifically, letting
$s\in\textrm{Par}$  where $\textrm{Par}\equiv\{r|r\in \{0,1\}^{n},\sum_i r_i\ge 2 \}$ is the set of $n$-bit
strings with at least two bits that are 1, no information about $x\cdot s=\bigoplus_{i}x_{i}s_{i}$ (termed the
$s$\emph{-parity}$)$ for any such $s$ can be transmitted to Bob (here $\oplus$ denotes sum modulo 2). \ This
task is similar to an $n$-to-$1$ quantum random access code~\cite{Galvao,Wiesner,ANTV,Hayashi} except that it
has a constraint of parity-obliviousness rather than a constraint on the potential information-carrying capacity
of the system used.

\begin{lemma}
\label{label:optimalclassicalprotocol}\ Classically, the optimal probability
of success in $n$-bit parity-oblivious multiplexing satisfies $p(b=x_{y}%
)\leq (n+1)/2n.$
\end{lemma}

\begin{proof} (For details, see Appendix A.) The only classical encodings of $x$ that
reveal no information about any parity (while encoding \emph{some} information about $x)$ are those that encode
only a single bit $x_{i}$ for some $i$. Given that $y$ is uniformly distributed,  it makes no difference which
bit it is.  Therefore, we may assume that Alice and Bob agree that Alice will always encode the first bit,
$x_1$. If $y=1,$ which occurs with probability $1/n$, then Bob can output $b=x_{y}$  and win.  With probability
$(n-1)/n,$ we have $y\neq1$ and in this case Bob can at best guess the value of $x_{y}$  and wins with
probability 1/2.
\end{proof}

What is the most general protocol that can be implemented in an arbitrary operational theory? \ For each input
string $x,$ Alice implements a preparation procedure $P_{x},$ and for each integer $y$, Bob implements a
binary-outcome measurement $M_{y},$ and
reports the outcome $b$ as his output. The probability of winning is%
\begin{equation}
p(b=x_{y})=\frac{1}{2^{n}n}\sum_{y\in\{1,...,n\}}\sum_{x\in\{0,1\}^{n}%
}p(b=x_{y}|P_{x},M_{y}) \label{eq:probsuccess}
\end{equation}
where $1/2^{n}n$ is the prior probability for a particular $x$ and
$y.$ The parity-oblivious constraint requires that for every
$s$-parity, there is no outcome of any measurement for which
posterior probabilities for $s$-parity $0$ and $s$-parity $1$ are
different, that is,
\begin{equation}
\forall s\forall M\forall k:\sum_{x|x\cdot s=0}p(P_{x}|k,M)=\sum_{x|x\cdot
s=1}p(P_{x}|k,M). \label{eq:POcondition}%
\end{equation}

\textbf{Noncontextuality inequality. }The main theoretical result of this letter is the following theorem.
\begin{theorem}
\label{thm:main} In an operational theory that admits a preparation
noncontextual hidden variable model, the optimal probability of
success in $n$-bit parity-oblivious multiplexing satisfies
$p(b=x_{y})\leq(n+1)/2n$.
\end{theorem}

\begin{proof}
 Define $P_{s,b}$ to be the procedure obtained by choosing
uniformly at random an $x$ such that $x\cdot s=b$ and implementing
$P_{x}$. Clearly, for any measurement $M,$ the probability of
outcome $k$ given preparation $P_{s,b}$ is simply
\begin{equation}
p(k|P_{s,b},M)=\frac{1}{2^{n-1}}\sum_{x|x\cdot s=b}p(k|P_{x},M).\label{eq:convexsum_operational}%
\end{equation}
Similarly, the probability of hidden variable $\lambda$ given an
implementation of $P_{s,b}$ is simply
\begin{equation}
p(\lambda|P_{s,b})=\frac{1}{2^{n-1}}\sum_{x|x\cdot s=b}p(\lambda
|P_{x}).\label{eq:convexsum_ontic}%
\end{equation}
  Now note that one can re-express the parity-oblivious condition, Eq.~(\ref{eq:POcondition}), as $\forall
s\forall M:\sum_{x|x\cdot s=0}p(k|P_{x},M)=\sum_{x|x\cdot s=1}p(k|P_{x},M)$  (it follows from Bayes' rule and
the uniformity of the prior over $x$).  Combining this with Eq.~(\ref{eq:convexsum_operational}), we infer that
 $\forall s\forall M:p(k|P_{s,0},M)=p(k|P_{s,1},M)$ which is simply the statement that mixed preparations
corresponding to opposite $s$-parities are indistinguishable by any measurement. \ But together with the
assumption that the hidden variable model is preparation noncontextual, Eq.~(\ref{eq:PNC}), this implies that
$\forall s:p(\lambda|P_{s,0})=p(\lambda|P_{s,1}),$ which states that mixed preparations corresponding to
opposite $s$-parities are also indistinguishable at the hidden variable level. Using
Eq.~(\ref{eq:convexsum_ontic}) and Bayes' rule again, we obtain
\begin{equation}
\forall s:\sum_{x|x\cdot s=0}p(P_{x}|\lambda)=\sum_{x|x\cdot s=1}%
p(P_{x}|\lambda).\label{eq:POhiddenvariable}%
\end{equation}
Therefore, even if one knew $\lambda$, the posterior probabilities for $s$-parity $0$ and $s$-parity $1$ would
be the same, that is, one would know nothing about any $s$-parity of $x$. \ The argument so far can be
summarized as follows: for preparation noncontextual models, parity-obliviousness at the operational level
implies parity-obliviousness at the level of the hidden variables. \

 The hidden state $\lambda$ provides a classical encoding of $x$.  But, as just shown, it is one that cannot
contain information about any $s$-parity.  We recall from lemma \ref{label:optimalclassicalprotocol} that such
encodings have information about at most one bit, $x_i$, of $x$.  Consequently, even if Bob could determine
$\lambda$ perfectly, he and Alice could at best achieve the optimal probability of success achievable in a
classical protocol (specified in lemma \ref{label:optimalclassicalprotocol}), while if Bob is limited in his
ability to determine $\lambda$ (as will be the case in general in a hidden variable model), they will do worse.
\end{proof}

\begin{figure}
[bt]
\begin{center}
\includegraphics[width=3.0in] {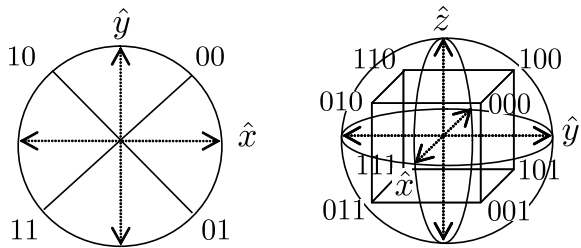}
\caption{Bloch representation of states and measurements in quantum $2$-bit and $3$-bit parity-oblivious
multiplexing.}\label{Blochspheres}
\end{center}
\end{figure}

\textbf{Quantum case. }We now consider how well one can achieve
parity-oblivious multiplexing in quantum theory. \ The following is
a protocol for the 2-bit case that uses a single qubit as the
quantum message. Alice encodes her 2 bits into the four pure quantum
states with Bloch vectors $(\pm\frac{1}{\sqrt{2}},\pm
\frac{1}{\sqrt{2}})$ equally distributed on an equatorial plane of
the Bloch sphere, as indicated in Fig.~\ref{Blochspheres} (recall
that a density operator $\rho$ is related to its Bloch vector
$\vec{r}$ by $\rho=\frac{1}{2}(I+\vec{r}\cdot \vec{\sigma}),$ where
$\vec{\sigma}$ is the vector of Pauli matrices). Bob measures along
the $\hat{x}$ axis if he wishes to learn the first bit, and along
the $\hat{y}$ axis if he wishes to learn the second. He guesses the
bit value $0$ upon obtaining the positive outcome. In all cases, the
guessed value is correct with probability
$\cos^{2}(\pi/8)\simeq\allowbreak0.853553.$ Meanwhile, no
information about the parity can be obtained by any quantum
measurement given that the parity 0 and parity 1 mixtures are
represented by the same density operator, $\frac{1}{2} \rho_{00}
+\frac{1}{2} \rho_{11} = \frac{1}{2} \rho_{01} + \frac{1}{2}
\rho_{10}=I/2.$  We have a violation of the NC inequality of
Thm.~\ref{thm:main}
because for $n=2$, the upper bound on the probability of success is
$3/4$.
By exploiting a connection with the Clauser-Horne-Shimony-Holt
inequality \cite{CHSH}, one can show that this protocol yields the
maximum possible quantum violation of the NC inequality.

A protocol for 3-bit parity-oblivious multiplexing using a single
qubit proceeds as follows. Alice encodes her three bits into a set
of eight pure quantum states associated with Bloch vectors
$(\pm\frac{1}{\sqrt{3}},\pm
\frac{1}{\sqrt{3}},\pm\frac{1}{\sqrt{3}})$ forming a cube inside the
Bloch sphere (see Fig.~\ref{Blochspheres}). \ Bob measures along the
$\hat{x},$ $\hat{y}$ or $\hat{z}$ axes to obtain the first, second
or third
bits. In all cases, the guessed value is correct with probability $\frac{1}{2}%
(1+\frac{1}{\sqrt{3}})\simeq\allowbreak0.788675.$ \ The mixture of the four states corresponding to $x_{1}\oplus
x_{2}=0$ (i.e. $s$-parity $0$ for $s=(1,1,0)$) is identical to the mixture of the four states corresponding to
$x_{1}\oplus x_{2}=1$ and is equal to $I/2$. \ Similarly for the two mixtures associated with each of the other
three parities, $x_{1}\oplus x_{3}$ ($s=(1,0,1)$)$,$ $x_{2}\oplus x_{3}$ ($s=(0,1,1)$)$,$ and $x_{1}\oplus
x_{2}\oplus x_{3}$ ($s=(1,1,1)$)$.$ The protocol is therefore parity-oblivious for all $s$-parities. Again we
have a violation of the NC inequality because for $n=3$ the upper bound on the probability of success is $2/3$.
It is an open question whether $0.788675$ is the maximum possible quantum violation.

The 2-bit protocol was originally presented as a 2-to-1 quantum
random access code by Wiesner \cite{Wiesner} and rediscovered in
Ref. \cite{ANTV}, while the 3-bit protocol was presented in Ref.
\cite{Hayashi} as an instance of a 3-to-1 quantum random access code
(the original idea is attributed to Chuang in Ref. \cite{ANTV}).

\textbf{Experimental results. } We experimentally demonstrate better-than-classical performance for $2$-bit and
$3$-bit parity-oblivious multiplexing
by implementing the quantum protocols using polarization qubits.
Photon pairs from downconversion
are coupled into single mode optical fibers. One photon acts as a trigger, while the other is used in the
experiment. Alice's state preparation consists of a fiber polarization controller, and a polarizing beam
displacer, rotated to the input state angle, used to ensure high-purity linearly polarized states for the
$2$-bit protocol. An additional quarter wave plate is used to prepare elliptically-polarized states for the
$3$-bit protocol. \ Bob's measurement consists of a polarizing beam displacer mounted in a computer-controlled
rotation mount, followed by a single photon counting module. For our demonstration, a detector is placed at only
a single output port of the beam displacer and the probability of each outcome is calculated from the relative
number of counts for a given beam displacer angle and the one orthogonal to it. (Further details of the
experimental set-up, including a figure, are provided in Appendix B.) Adjustment of the beam displacer and
quarter wave plate angles allows measurement of the horizontal/vertical basis, the diagonal/anti-diagonal basis
and the right/left-circular basis.
Valid measurement events are heralded by a coincidence count between the directly detected photon and the
experiment photon. These experimental procedures for a given $x$ and $y$ define the preparation $P_{x}$ and the
measurement $M_{y}$ respectively.

We obtained probabilities $p(k=x_{y}|P_{x},M_{y})$ by accumulating
statistics over approximately $3.5\times10^{7}$ coincidence counts
for each $x$ and $y$ in the 2-bit scheme and
$2.\,\allowbreak4\times10^{7}$ in the 3-bit scheme. Using
Eq.~(\ref{eq:probsuccess}), we calculated the 2-bit and 3-bit
probabilities of success to be $p(b=x_{y})=0.851929\pm 0.000030$ and
$p(b=x_{y})=0.786476\pm 0.000017$ respectively. The errors were
determined from the Poissonian counting statistics of the parametric
source and the small repeatability error in the wave plate settings,
using standard error analysis techniques. These probabilities of
success violate the NC inequality of Thm.~\ref{thm:main} with a high
degree of confidence: $3410$ and $6922$ standard deviations
respectively. \ They are also close to the predicted quantum values
of $0.853553$ and $0.788675$, achieving a violation that is $98.4\%$
and $98.2\%$ respectively of the gap between the NC bound and the
quantum value.

Just as Bell inequality violations are only surprising given the absence of signalling between the two wings of
the experiment, the NC inequality violations are only surprising given the parity-oblivious property. However,
whereas one can establish the absence of signalling by confirming that the two wings are space-like separated,
one must directly test for transmission of information about the parity in our experiment. A consideration of
how this is to be accomplished highlights two shortcomings in the operational definition of preparation
noncontextuality of Eq.~(\ref{eq:PNC}):
in practice one can never implement \emph{all} measurements and one never finds truly \emph{identical}
statistics. The first issue may be addressed by relying on previous experimental evidence for the existence of a
tomographically complete set of measurements -- one from which the statistics of any other measurement can be
calculated -- and testing indistinguishability relative to this set alone, as we shall do here.  The second
issue may be addressed by presuming a kind of continuity: closeness of experimental statistics implies closeness
of the representations in the model \cite{Spe05} (this parallels the problem of dealing with imperfect alignment
in traditional proofs of contextuality \cite{Meyer}, where continuity also provides an answer
\cite{previouswork,replytoMeyer}). In the present work, we simply demonstrate that the experimental statistics
are \emph{close} to parity-oblivious while yielding a large violation of the noncontextuality inequalities, and
leave a more detailed analysis for future work.


We quantify the obliviousness of our experimental protocol for a particular $s$-parity by the maximum
probability that Bob can correctly estimate this parity in a variation over all measurements.  One can estimate
this by implementing a tomographically complete set of measurements, then reconstructing the states $\rho_0$ and
$\rho_1$ associated with $s$-parity 0 and $s$-parity 1, and finally making use of the fact that the maximum
probability of discriminating these states is $\frac{1}{2}+\frac{1} {4}\mathrm{Tr}|\rho_0-\rho_1|.$ Among all
$s$-parities, we calculate the largest such probability to be $0.5020 \pm 0.0002$. This calculation is not
sufficient, however, because it neglects an imperfection in the experiment that also contributes to leakage of
information about the parity, namely, that there is a small probability of more than one photon being sent to
the experiment. By our characterization of the source, we estimate the probability of two photons to be
$0.007\pm 0.003$ relative to the single photon generation probability.
 If two photons pass through the polarizers in the ideal protocol, the maximum
probability of correctly estimating the parity can be quite far from 1/2: it is 3/4 in the case of the 2-bit
scheme and 2/3 for three of the four $s$-parities in the 3-bit scheme. However, the fact that this possibility
occurs with low probability implies that the two-photon contribution to the probability of correct estimation is
comparable to the one-photon contribution. (Contributions from three or more photons are negligible in
comparison). The weighted average of these contributions is easily calculated and the largest, among all
$s$-parities, is found to be $0.504 \pm 0.002$. The fact that this is within one percent of 1/2 demonstrates
that our experimental protocols are indeed close to parity-oblivious.



Given that the quantum protocols described herein are also 2-to-1
and 3-to-1 random access codes, our results constitute the first
experimental demonstration of a quantum advantage for these tasks as
well.

Finally, it is worth noting that every Bell inequality is a special case of an NC inequality where all
assumptions of noncontextuality are justified by locality \cite{Spe05}.\ \ Consequently, every experimental
violation of a Bell inequality demonstrates the impossibility of a noncontextual hidden variable model. Indeed,
this is \emph{all} that can be demonstrated by those that fail to seal the locality loophole \cite{Row01,Has03}.
 Nonetheless, a dedicated experiment of the sort we have described here can achieve a large violation with
high confidence at a smaller cost of experimental effort.


\textbf{\newline Acknowledgements. }R.W.S. thanks M.~Leifer and J.~Barrett for helpful discussions. This work
has been supported by the Australian Research Council, an IARPA-funded US Army Research Office contract, NWO
VICI project 639-023-302, the Dutch BSIK/BRICKS project, the EU's FP6-FET Integrated Projects SCALA (CT-015714)
and QAP (CT-015848), and the Royal Society.


\section{Appendix A: Optimal classical protocol for n-bit parity-oblivious multiplexing}
We here provide a more detailed proof of lemma~1.
First, note that by the assumption of parity-obliviousness, the
classical message $m$ sent from Alice to Bob must satisfy%
\begin{equation}\label{eq:POclassical}
\forall s:\sum_{x|x\cdot s=0}p(P_{x}|m)=\sum_{x|x\cdot s=1}p(P_{x}|m)
\end{equation}
By Bayes' theorem and the fact that the distribution over inputs $x$ is uniform, we can rewrite this as a
constraint on $p(m|P_{x}),$ namely,
\begin{equation}
\forall s:\sum_{x|x\cdot s=0}p(m|P_{x})=\sum_{x|x\cdot s=1}p(m|P_{x}%
).\label{eq:POclassical2}%
\end{equation}
As we will demonstrate (at the end of this section), this implies that $p(m|P_{x})$ has the form
\begin{align}
p(m|P_x) & = p(0)p_{0}(m) \nonumber \\
& +\sum_{i=1}^{n}p(i)\left[  p_{i,0}(m)\delta_{x_{i},0} +p_{i,1}(m)\delta_{x_{i},1}\right],
\label{eq:robisgreat}
\end{align}
where $p(i)$ is a normalized probability distribution on $\{0,\dots,n\},$ the functions $p_{0}(m),p_{i,0}(m)$
and $p_{i,1}(m)$ are normalized probability distributions over $m$, and where $\delta_{a,b}$ is the Kronecker
delta function (equal to $1$ if $a=b$ and $0$ otherwise).

It follows that any classical parity-oblivious multiplexing protocol can be interpreted as follows: Alice
generates an integer $i\in\{0,\dots,n\}$ from the distribution $p(i).$ \ Upon obtaining $i=0,$ she sends a
message $m$ chosen from the distribution $p_{0}(m)$ (independent of the value of $x).$ \ Upon obtaining
$i\in\{1,\dots,n\},$ she sends a message $m$ chosen from one of two distributions, depending on the value of the
$i$th bit of $x:$ the distribution is $p_{i,0}(m)$ if $x_{i}=0$ and $p_{i,1}(m)$ if $x_{i}=1.$

We now determine the choice of these distributions that leads to a maximum probability of winning. \ First note
that if $i=0$, Bob gets no information about $x.$ \ This is clearly not optimal, so we may set $p(0)=0.$ \ Next
note that the amount that Bob learns about $x_{i}$ depends on his ability to distinguish $p_{i,0}(m)$ from
$p_{i,1}(m).$ To optimize the amount that Bob can learn, $p_{i,0}(m)$ and $p_{i,1}(m)$ must be chosen to be
perfectly distinguishable. \ This is only possible if they are completely non-overlapping, that is, if
$p_{i,0}(m)p_{i,1}(m)=0$.

In an optimal decoding, \strut Bob simply determines whether $m$ is in the support of $p_{y,0}(m)$ or of
$p_{y,1}(m)$ and outputs $b=0$ or $1$ accordingly. \ This is optimal for the following reason. The message $m$
only contains information about $x_{y}$ if Alice happened to generate an $i$ that coincides with $y$ and in this
case Bob will output $b=x_{y}$ with probability $1.$\ When $i$ does not coincide with $y,$ Bob gets no
information about $x_{y}$ from $m,$ so it is irrelevant what he outputs; given that $x_{y}$ is equally likely to
be $0$ or $1,$ his probability of having generated the correct output will be $1/2$.

Finally, given that $y$ is chosen uniformly at random, the probability of $i$ coinciding with $y$ is $1/n$, so
that the overall probability of a correct output is $\frac{1}{n}(1)+\left( 1-\frac{1}{n}\right)  (\frac{1}{2})$
$= (n+1)/2n.$

It is worth noting that there are many natural schemes that achieve the optimum:
\begin{itemize}
\item If $p(i)=\delta_{i,j}$ for some particular $j\in\{1,\dots,n\}$,
and $p_{j,0}(m)p_{j,1}(m)=0,$ then Alice has simply encoded the value of $x_{j}$ in her message.
\item  For $p(i)$ an arbitrary distribution over $\{1,\dots,n\},$ if $p_{i,0}%
(m)p_{i,1}(m)=0$ for all $i,$ then Alice has simply chosen a value $i$ $\in\{1,\dots,n\}$ according to this
distribution and encoded $x_{i}$ in her message.
\item For $p(i)$ an arbitrary distribution over $\{1,\dots,n\},$ if $p_{i,b}%
(m)p_{i^{\prime},b^{\prime}}(m)=0$ when either $b\neq b^{\prime}$ or $i\neq i^{\prime},$ then Alice has encoded
both $i$ and $x_{i}$ in her message.
\end{itemize}

It remains to prove our claim that the parity-oblivious constraint, Eq.~(\ref{eq:POclassical2}), implies the
decomposition of $p(m|P_{x})$ described in Eq.~(\ref{eq:robisgreat}). We do this using Fourier analysis over
$\mathbb{Z}_{2}^{n}$.
Let $r\in\{0,1\}^{n}$. Define functions $\chi_{r}%
\strut:\{0,1\}^{n}\rightarrow\lbrack-1,1]$ where
\[
\chi_{r}\strut(x):=(-1)^{x\cdot r}.
\]
These form an orthonormal set because
\[
\sum_{x}\chi_{r}\strut(x)\chi_{r^{\prime}}\strut(x)=\sum_{x\in\{0,1\}^{n}%
}(-1)^{x\cdot(r\oplus r^{\prime})}=2^{n}\delta_{r,r^{\prime}}.
\]
Moreover, noting that the dimensionality of the space of functions on $\{0,1\}^{n}$ is $2^{n}$ (a parameter for
every input string) and that there are $2^{n}$ values of $r$, we see that the $\chi_{r}$ form an orthonormal
basis of the function space. It follows that we can write $p(m|P_x)$ in the Fourier series
\[
p(m|P_x)=\sum_{r}\hat{p}(m,r)\chi_{r}(x).
\]
We infer that
\begin{align*}
2^{n}\hat{p}(m,r)&=\sum_{x}\chi_{r}(x)p(m|P_x) \\& =\sum_{x|x\cdot r=0}p(m|P_x)-\sum_{x|x\cdot r=1}p(m|P_x).
\end{align*}
Combining this with the parity-obliviousness condition, Eq.~(\ref{eq:POclassical2}), one obtains
\[
\forall s\in \text{Par} : \hat{p}(m,s)=0.
\]
Consequently, the only strings $r$ for which $\hat {p}(m,r)\neq0$ are those with Hamming weight $0$ or $1$.
Denoting the Fourier coefficients of the all zero string by $\hat{p}_{0}(m)$ and that of the string with a
single 1 at position $i$ by $\hat{p}_{i}(m)$, we have
\[
p(m|P_x)=\hat{p}_{0}(m)+\sum_{i=1}^{n}\hat{p}_{i}(m)(-1)^{x_{i}}.
\]
Because $(-1)^{x_{i}}=\delta_{x_i,0}-\delta_{x_i,1}$ and $1 = \delta_{x_i,0}+\delta_{x_i,1}$, we can write%
\begin{align}\label{eq:benisgroovy}
p(m|P_x)=a_{0}(m)+\sum_{i=1}^{n}\left[a_{i,0}(m)\delta_{x_i,0}+a_{i,1}(m)\delta_{x_i,1}\right] 
\end{align}
where we have defined nonnegative coefficients%
\begin{align*}
  a_{i,0}(m) = 2\hat p_i(m),\ a_{i,1}(m) = 0 \qquad& \text{if $\sgn(\hat{p}_{i}(m))\geq0$},\\
  a_{i,0}(m) = 0,\ a_{i,1}(m) = -2\hat p_i(m) \qquad& \text{if $\sgn(\hat{p}_{i}(m))<0$};
\end{align*}
and we have implicitly defined a constant $a_{0}(m)$, which we presently show is also nonnegative. To do this,
we define an $n$-bit string $z(m)$ that encodes the
signs of the Fourier coefficients. Specifically, $z(m)$ is defined by%
\[
z_{i}(m)\equiv%
\begin{cases}
1 & \text{if $\sgn(\hat{p}_{i}(m))\geq0$},\\
0 & \text{if $\sgn(\hat{p}_{i}(m))<0.$}%
\end{cases}
\]

It follows from this definition that
\begin{align*}
  a_{i,0}(m)\delta_{z_i(m),0}+a_{i,1}(m)\delta_{z_i(m),1} = 0
\end{align*}
for all $i$,
and consequently that%
\begin{align*}
p(m|P_{z(m)})  & =a_{0}(m),
\end{align*}
which establishes that $a_{0}(m)\geq 0$.

Finally, we show that Eq.~(\ref{eq:benisgroovy}) can be put into the form of Eq.~\eqref{eq:robisgreat}.
By the normalization of the distribution $p(m|P_x),$ we have%
\[
1=\sum_{m}p(m|P_x)=\sum_{m}a_{0}(m)+\sum_{i=1}^{n}\sum_{m}a_{i,x_{i}}(m),
\]
for all $x.$ Defining $A_{0}=\sum_{m}a_{0}(m)$ and $A_{i,x_i}=\sum_{m}%
a_{i,x_i}(m),$ we have
\[
A_{0}+\sum_{i=1}^{n}A_{i,x_{i}}=1,
\]
for all $x,$ which implies that $\sum_{i=1}^{n}A_{i,x_{i}}$ is independent of $x$ and in particular of $x_i$.
We deduce that
\[
A_{i,0}=A_{i,1}
\]
for all $i$. Eq.~\eqref{eq:robisgreat} now follows from
Eq.~(\ref{eq:benisgroovy}) by identifying%
\begin{align*}
p(0)  & =A_{0}\\
p(i)  & =A_{i,0}=A_{i,1}%
\end{align*}
and
\begin{align*}
p_{0}(m)  & =a_{0}(m)/p(0)\\
p_{i,b}(m)  & =a_{i,b}(m)/p(i).
\end{align*}
if $p(0), p(i) \neq 0$.

\section{Appendix B: Experimental details}
\begin{figure}
[pbt]
\begin{center}
\includegraphics[width=3.0in] {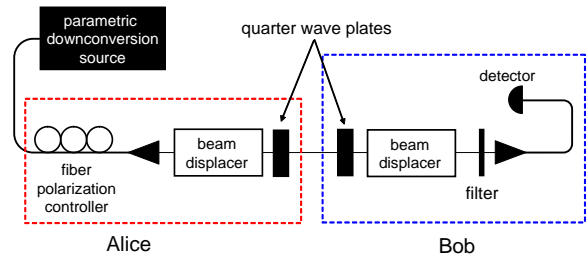}
\caption{Experimental set-up.  The parametric downconversion source provides single photons to the experiment.
Detection events in the experiment are counted in coincidence with the downconversion trigger photon (not
shown).}  \label{setup}
\end{center}
\end{figure}

A schematic of the experimental set-up is provided in
Fig.~\ref{setup}. We used type-I downconversion in bismuth borate (BiBO) to generate pairs of $820$ nm,
horizontally polarized single photons from a 410 nm, 60 mW continuous-wave diode laser. A 10 nm FWHM
interference filter is used to reject background light. In the experiment, we obtained coincidence rates (2.5 ns
window) of approximately 23100 pairs/s in the 2-bit scheme and 15200 pairs/s in the 3-bit scheme.

Although we chose to implement the experiment with a heralded mode of a downconversion source, similar results
could also have been obtained with weak coherent states (using the same measurements) \footnote{The Wigner
representation would not provide a classical statistical model of the experiment because the representation of
the measurements would be nonpositive.}.  In both cases, one must postselect on not finding the vacuum
(implying, incidentally, that the detector loophole is not sealed~\cite{detectorloophole}), and in both cases
there is a small amplitude for more than one photon and hence a small amount of leaked parity information. Our
choice was motivated by differences in ideal performance -- it is only for the downconversion scheme that the
leakage of parity information can be eliminated in principle (through the use of true single photons heralded by
efficient number-resolving detectors). Nonetheless, this ideal has not yet been realized.



\begin{thebibliography}{99}                                                                                               %

\bibitem {BKS}J.\ S.\ Bell, Rev. Mod. Phys. \textbf{38}, 447 (1966);
S.\ Kochen and E.\ P.\ Specker, J. Math. Mech. \textbf{17}, 59 (1967).

\bibitem {Spe05}R. W. Spekkens, Phys. Rev. A \textbf{71}, 052108 (2005).

\bibitem {Bell64}J.\ S.\ Bell, Physics \textbf{1}, 195 (1964).

\bibitem{previouswork} A.\ Cabello and G.\ Garcia-Alcaine, Phys. Rev. Lett.
\textbf{80,} 1797 (1998); C.\ Simon, C.\ Brukner and A.\ Zeilinger,
Phys. Rev. Lett. \textbf{86,} 4427 (2001); J.-A.\ Larsson, Europhys.
Lett. \textbf{58}, 799 (2002).

\bibitem {Galvao}E. F. Galv\~{a}o, Ph.D. thesis, arXiv:quant-ph/0212124v1.

\bibitem {Wiesner}S. Wiesner, Sigact News \textbf{15}, 78 (1983).

\bibitem {ANTV}A. Ambainis, A. Nayak, A. Ta-Shma, and U. Vazirani, in
Proceedings of the 31st Annual ACM Symposium on the Theory of Computing (ACM Press, New York, 1999).

\bibitem {Hayashi}M. Hayashi \textit{et al.}, New Journal of Physics \textbf{8}, 129 (2006).

\bibitem {CHSH}J. F. Clauser, M.A. Horne, A. Shimony and R. A. Holt, Phys.
Rev. Lett. \textbf{23}, 880 (1969).

\bibitem{Row01} M. A. Rowe \textit{et al.}, Nature \textbf{409}, 791 (2001).
\bibitem{Has03} Y.~Hasegawa \textit{et al.}, Nature \textbf{425}, 45 (2003).



\bibitem {Meyer}D. A. Meyer, Phys. Rev. Lett. \textbf{83} 3751 (1999).

\bibitem {replytoMeyer}N. D. Mermin, arXiv:quant-ph/9912081.


\bibitem{detectorloophole} P. Pearle, Phys. Rev. D, \textbf{2}, 1418 (1970).


\end{thebibliography}

\end{document}